\documentclass[12pt]{article}
\usepackage{epsfig}
\newcommand{\lsim}{
\mathrel{\hbox{\rlap{\hbox{\lower4pt\hbox{$\sim$}}}\hbox{$<$}}}}
\newcommand{\gsim}{
\mathrel{\hbox{\rlap{\hbox{\lower4pt\hbox{$\sim$}}}\hbox{$>$}}}}
\begin{document}
%
%
\newcommand{\x}{\cdot}
\newcommand{\ra}{\rightarrow}
\begin{titlepage}
\vspace{3 ex}
%
%
\vspace*{0.2truecm}
\begin{flushright}
\begin{tabular}{l}
DESY 02--110\\
hep--ph/0208083\\
August 2002
\end{tabular}
\end{flushright}

\vspace*{1.0truecm}

\begin{center}
{
\LARGE \bf \rule{0mm}{7mm}{\boldmath  Status and Prospects of CKM 
Phase Determinations}\\
}

\vspace{4ex}

\vspace*{1.3truecm}

{\large
Robert Fleischer\\
}
\vspace{1 ex}

{\em
Deutsches Elektronen-Synchrotron DESY, D--22607 Hamburg, Germany \\
}
\vspace{2 ex}
%
%
\end{center}

\vspace{2 ex}
%
%

\vspace*{0.7truecm}

\begin{abstract}
We give an overview of direct determinations of the angles of the unitarity 
triangle of the CKM matrix, using CP-violating effects in $B$-meson decays. 
After a discussion of $B\to\pi K$ modes, which can be described efficiently 
through allowed regions in observable space and play an important r\^ole to 
determine $\gamma$, we turn to extractions of the $B^0_d$--$\overline{B^0_d}$ 
mixing phase $\phi_d$, which equals $2\beta$ in the Standard Model, from 
$B_d\to J/\psi K_{\rm S}$, and emphasize that it is important to determine 
this phase unambiguously. Finally, we focus on $B_d\to\pi^+\pi^-$, where 
recent $B$-factory data point towards large penguin contributions. The 
question arises now how the CP-violating observables of this mode can be 
transformed into information on the angles of the unitarity triangle. A 
promising tool to achieve this goal is offered by $B_s\to K^+K^-$, which is 
very accessible at hadronic $B$ experiments, and allows a determination 
of $\phi_d$ and $\gamma$. A variant for the $e^+e^-$ $B$-factories is 
provided by $B_d\to\pi^\mp K^\pm$, where data are already available, 
pointing to an exciting picture and a highly constrained allowed region in 
$B_s\to K^+K^-$ observable space. 
\end{abstract}

\vspace*{1.3truecm}

\begin{center}
{\sl Invited talk at the\\
8th International Conference on
$B$ Physics at Hadron Machines -- BEAUTY2002\\ 
Santiago de Compostela, Spain, 17--21 June 2002\\ 
To appear in the Proceedings}
\end{center}

\end{titlepage}

%
\setlength{\oddsidemargin}{0 cm}
\setlength{\evensidemargin}{0 cm}
\setlength{\topmargin}{0.5 cm}
\setlength{\textheight}{22 cm}
\setlength{\textwidth}{16 cm}
\setcounter{totalnumber}{20}

\clearpage\mbox{}\clearpage

\pagestyle{plain}
\setcounter{page}{1}
%

%
%
%
\section{Introduction}\label{sec:intro}
The non-conservation of the CP symmetry in weak interactions is a mystery 
since its discovery through $K_{\rm L}\to\pi\pi$ decays in 1964 \cite{CP-obs}.
Before the start of the $B$ factories, CP-violating effects could only be
studied in the kaon system. In this decade, decays of neutral and charged 
$B$-mesons will provide further valueable insights into CP violation, in 
particular stringent tests of the Kobayashi--Maskawa mechanism of CP 
violation \cite{KM}, which introduces this exciting phenomenon to the 
Standard Model of electroweak interactions. Using the ``gold-plated'' mode 
$B_d\to J/\psi K_{\rm S}$ \cite{bisa}, large CP-violating effects could 
actually be established in the $B$ system in 2001 
\cite{BaBar-CP-obs,Belle-CP-obs}. The central target of the $B$ 
factories is the famous unitarity triangle of the Cabibbo--Kobayashi--Maskawa 
(CKM) matrix, with the goal to overconstrain this triangle as much as 
possible through independent measurements of its sides and angles. A detailed 
review of this topic can be found in \cite{RF-Phys-Rep}.

\begin{figure}[h]
\centerline{
\epsfysize=6.0truecm
{\epsffile{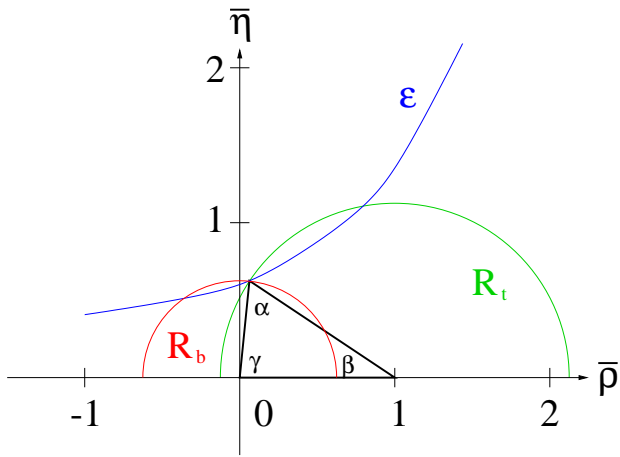}}}
\caption{The ``standard analysis'' to determine the unitarity triangle in 
the $\overline{\rho}$--$\overline{\eta}$ plane.}\label{fig:UT-det}
\end{figure}

The ``standard analysis'' to determine the apex of the unitarity triangle 
in the plane of the generalized Wolfenstein parameters $\overline{\rho}$ 
and $\overline{\eta}$ \cite{wolf,BLO} employs the following ingredients, 
as illustrated in Fig.~\ref{fig:UT-det}:
\begin{itemize}
\item Exclusive and inclusive semi-leptonic $B$ decays caused by
$b\to c\ell\overline{\nu}_{\ell}, u\ell\overline{\nu}_{\ell}$ quark-level
transitions \cite{ligeti}, fixing a circle with radius $R_b$ around $(0,0)$.
\item $B^0_q$--$\overline{B^0_q}$ mixing ($q\in\{d,s\}$), fixing a circle 
with radius $R_t$ around $(1,0)$.
\item Indirect CP violation in the neutral kaon system, $\varepsilon$, fixing
a hyperbola.
\end{itemize}
Many different strategies to deal with the corresponding theoretical and 
experimental uncertainties can be found in the literature. The most 
important ones are the simple scanning approach \cite{scanning}, 
the Gaussian approach \cite{gauss}, the BaBar 95\% scanning method
\cite{babar-scan}, the Bayesian approach \cite{bayesian}, and the
statistical approach developed in \cite{hoeck}. A detailed discussion of 
these approaches is beyond the scope of this presentation. Let us here just
give typical ranges for $\alpha$, $\beta$ and $\gamma$ that are implied by 
these strategies:
\begin{equation}\label{UT-Fit-ranges}
70^\circ\lsim\alpha\lsim 130^\circ, \quad
15^\circ\lsim\beta\lsim 35^\circ, \quad
50^\circ\lsim\gamma\lsim 70^\circ.
\end{equation}

As far as the range for $\gamma$ is concerned, a particularly important
constraint is provided by lower bounds on the $B_s$ mass difference 
$\Delta M_s$, which imply upper bounds on the side $R_t$ of the unitarity
triangle \cite{Buras-ICHEP96}:
\begin{equation}\label{DMs-constr}
\left(R_t\right)_{\rm max}=0.83\times\xi\times\sqrt{\frac{15.0\,
{\rm ps}^{-1}}{\left(\Delta M_s\right)_{\rm min}}},
\end{equation}
where 
\begin{equation}\label{xi-def}
\xi\equiv\frac{\sqrt{\hat B_{B_s}}f_{B_s}}{\sqrt{\hat B_{B_d}}f_{B_d}}=
1.15\pm0.06
\end{equation}
measures $SU(3)$-breaking effects in non-perturbative mixing and decay
parameters. The strong experimental bound $\Delta M_s>14.6\,{\rm ps}^{-1}$ 
(95\% C.L.) \cite{rozen} excludes already a large part in the 
$\overline{\rho}$--$\overline{\eta}$ plane, implying in particular 
$\gamma<90^\circ$. In a recent paper \cite{KroRy}, it is argued that $\xi$ 
may actually be significantly larger than the conventional range given in 
(\ref{xi-def}), $\xi=1.32\pm0.10$. In this case, the excluded range in
the $\overline{\rho}$--$\overline{\eta}$ plane would be reduced, shifting
the upper limit for $\gamma$ closer to $90^\circ$. Hopefully, the status of
$\xi$ will be clarified soon. In the near future, run II of the Tevatron 
should provide a measurement of $\Delta M_s$ \cite{Paulini,Kouznetsov}, 
thereby constraining the unitarity triangle and $\gamma$ in a much more 
stringent way. 

After the discovery of CP violation in the $B$ system, the major goal is 
now the {\it direct} determination of the angles of the unitarity triangle
through CP-B measurements. To this end, various approaches were proposed
over recent years. Here we shall have a closer look at the following
strategies: in Section~\ref{sec:BpiK}, we discuss $B\to\pi K$ modes, which
are promising to extract $\gamma$. In Section~\ref{sec:phid}, we turn to 
the determination of the $B^0_d$--$\overline{B^0_d}$ mixing phase 
$\phi_d$, which equals $2\beta$ in the Standard Model, from the ``gold-plated''
decay $B_d\to J/\psi K_{\rm S}$, and emphasize that it is important to 
determine this phase unambiguously. Another particularly interesting mode, 
the $B_d\to\pi^+\pi^-$ channel, is the subject of Section~\ref{sec:Bpipi}. 
Finally, concluding remarks and a brief outlook are given in 
Section~\ref{sec:concl}.

\boldmath
\section{Extracting $\gamma$ from $B\to\pi K$ Decays}\label{sec:BpiK}
\unboldmath
If we employ flavour-symmetry arguments and make plausible dynamical 
assumptions, $B\to\pi K$ deacys allow determinations of $\gamma$ and 
hadronic parameters with a ``minimal'' theoretical input 
\cite{GRL}--\cite{GR-BpiK-recent}. Alternative approaches, relying on 
a more extensive use of theory, are provided by the recently developed 
``QCD factorization'' \cite{QCDF} and ``PQCD'' \cite{PQCD} approaches, 
which allow furthermore a reduction of the theoretical uncertainties 
of the flavour-symmetry strategies discussed here. 

\subsection{General Features and Observables}
As can be seen by looking at the corresponding Feynman diagrams, $B\to\pi K$
modes may receive contributions from penguin and tree-diagram-like topologies,
where the latter bring the CKM angle $\gamma$ into the game. Because of the 
small CKM factor $|V_{us}V_{ub}^\ast/(V_{ts}V_{tb}^\ast)|\approx0.02$, the 
QCD penguin topologies play actually the dominant r\^ole, despite their loop 
suppression. As far as electroweak (EW) penguins are concerned, they 
contribute in colour-suppressed form to $B_d^0\to\pi^-K^+$ and 
$B^+\to\pi^+ K^0$, and are hence expected to play a minor r\^ole in these 
modes. On the other hand, EW penguins may also contribute in colour-allowed 
form to $B^+\to\pi^0K^+$ and $B^0_d\to\pi^0K^0$, and may here compete with 
tree-diagram-like topologies. 

\begin{table}
\centering
\begin{tabular}{|c||c|c|c|c|}
\hline
Observable & CLEO   & BaBar   & Belle & Average\\
\hline
$R$ & $1.00\pm0.30$ & $1.08\pm0.16$ & $1.23\pm0.26$ & $1.10\pm0.14$\\
$R_{\rm c}$ & $1.27\pm0.47$ & $1.27\pm0.24$ & $1.33\pm0.37$ & $1.29\pm0.21$\\
$R_{\rm n}$ & $0.59\pm0.27$ & $1.09\pm0.42$ & $1.42\pm0.68$ & $1.03\pm0.28$\\
\hline
\end{tabular}
\caption{CP-conserving $B\to \pi K$ observables as defined in
(\ref{mixed-obs})--(\ref{neut-obs}). For the evaluation of $R$, we have
used $\tau_{B^+}/\tau_{B^0_d}=1.060\pm0.029$. The data refer to
\cite{CLEO-BpiK,bartoldus}.}\label{tab:BPIK-obs}
\end{table}

\begin{table}
\centering
\begin{tabular}{|c||c|c|c|c|}
\hline
Observable & CLEO & BaBar  & Belle & Average\\
\hline
$A_0$ & $0.04\pm0.16$ & $0.05\pm0.07$ & $0.07\pm0.09$ & $0.05\pm0.07$\\
$A_0^{\rm c}$ & $0.37\pm0.32$ & $0.00\pm0.23$ & $0.05\pm0.26$ & $0.14\pm0.16$\\
$A_0^{\rm n}$ & $0.02\pm0.10$ & $0.05\pm0.07$ & $0.09\pm0.13$ & $0.05\pm0.06$\\
\hline
\end{tabular}
\caption{CP-violating $B\to \pi K$ observables as defined in
(\ref{mixed-obs})--(\ref{neut-obs}). For the evaluation
of $A_0$, we have used $\tau_{B^+}/\tau_{B^0_d}=1.060\pm0.029$.
The data refer to \cite{CLEO-BpiK-CPV,bartoldus}.}\label{tab:BPIK-obs-CPV}
\end{table}

Relations between the $B\to\pi K$ amplitudes that are implied by the $SU(2)$
isospin flavour symmetry of strong interactions suggest the following 
combinations to probe $\gamma$: the ``mixed'' $B^\pm\to\pi^\pm K$, 
$B_d\to\pi^\mp K^\pm$ system \cite{PAPIII}--\cite{defan}, the ``charged'' 
$B^\pm\to\pi^\pm K$, $B^\pm\to\pi^0K^\pm$ system \cite{NR}--\cite{BF-neutral1},
and the ``neutral'' $B_d\to\pi^0 K$, $B_d\to\pi^\mp K^\pm$ system 
\cite{BF-neutral1,BF-neutral2}. Correspondingly, we may introduce the
following sets of observables \cite{BF-neutral1}:
\begin{equation}\label{mixed-obs}
\mbox{}\hspace*{0.4truecm}\left\{\begin{array}{c}R\\A_0\end{array}\right\}
\equiv\left[\frac{\mbox{BR}(B^0_d\to\pi^-K^+)\pm
\mbox{BR}(\overline{B^0_d}\to\pi^+K^-)}{\mbox{BR}(B^+\to\pi^+K^0)+
\mbox{BR}(B^-\to\pi^-\overline{K^0})}\right]\frac{\tau_{B^+}}{\tau_{B^0_d}}
\end{equation}
\begin{equation}\label{charged-obs}
\left\{\begin{array}{c}R_{\rm c}\\A_0^{\rm c}\end{array}\right\}
\equiv2\left[\frac{\mbox{BR}(B^+\to\pi^0K^+)\pm
\mbox{BR}(B^-\to\pi^0K^-)}{\mbox{BR}(B^+\to\pi^+K^0)+
\mbox{BR}(B^-\to\pi^-\overline{K^0})}\right]
\end{equation}
\begin{equation}\label{neut-obs}
\left\{\begin{array}{c}R_{\rm n}\\A_0^{\rm n}\end{array}\right\}
\equiv\frac{1}{2}\left[\frac{\mbox{BR}(B^0_d\to\pi^-K^+)\pm
\mbox{BR}(\overline{B^0_d}\to\pi^+K^-)}{\mbox{BR}(B^0_d\to\pi^0K^0)+
\mbox{BR}(\overline{B^0_d}\to\pi^0\overline{K^0})}\right].
\end{equation}
The experimental status of these observables is summarized in 
Tables~\ref{tab:BPIK-obs} and \ref{tab:BPIK-obs-CPV}, where the CLEO
numbers refer to \cite{CLEO-BpiK,CLEO-BpiK-CPV}, and those of BaBar and
Belle to the recent updates reviewed in \cite{bartoldus}. Moreover, there 
are stringent constraints on CP violation in $B^\pm\to\pi^\pm K$:
\begin{equation}\label{ACP-Bc}
{\cal A}_{\rm CP}(B^\pm\to\pi^\pm K)=\left\{\begin{array}{ll}
0.17\pm0.10\pm0.02 & \mbox{(BaBar)}\\ 
-0.46\pm0.15\pm0.02 & \mbox{(Belle).}
\end{array}\right.
\end{equation}
Let us note that a very recent preliminary study of Belle indicates that
the large asymmetry in (\ref{ACP-Bc}) is due to a $3 \sigma$ fluctuation
\cite{suzuki}. Within the Standard Model, a sizeable value of 
${\cal A}_{\rm CP}(B^\pm\to\pi^\pm K)$ could be induced by large 
rescattering effects. Other important indicators for
such processes are branching ratios for $B\to KK$ decays, which are already 
strongly contstrained by the $B$ factories, and would allow us to take
into account rescattering effects in the extraction of $\gamma$ from
$B\to\pi K$ modes \cite{defan,neubert,BF-neutral1,FSI-incl}. Let us note 
that also the QCD factorization approach \cite{QCDF} is not in favour of large
rescattering processes. For simplicity, we shall neglect such effects in 
the discussion given below. 

Interestingly, already CP-averaged $B\to\pi K$ branching ratios may lead 
to non-trivial constraints on $\gamma$ \cite{FM,NR,BF-neutral1}, provided 
the corresponding $R_{\rm (c,n)}$ observables are found to be sufficiently 
different from one. The final goal is, however, to determine $\gamma$. Let 
us first turn to the charged and neutral $B\to\pi K$ systems in the following 
subsection.

\boldmath
\subsection{The Charged and Neutral $B\to\pi K$ Systems}\label{BpiK-cn}
\unboldmath
The starting point of our considerations are relations between the charged
and neutral $B\to\pi K$ amplitudes that follow from the $SU(2)$ isospin
symmetry of strong interactions. Assuming moreover that the rescattering
effects discussed above are small, we arrive at a parametrization of the
following structure \cite{BF-neutral1} (for an alternative one, see
\cite{neubert}):
\begin{eqnarray}
R_{\rm c,n}&=&1-2r_{\rm c,n}\left(\cos\gamma-q\right)\cos\delta_{\rm c,n}
+\left(1-2q\cos\gamma+q^2\right)r_{\rm c,n}^2\label{Rcn-par}\\
A_0^{\rm c,n}&=&2r_{\rm c,n}\sin\delta_{\rm c,n}\sin\gamma.\label{Acn-par}
\end{eqnarray}
Here $r_{\rm c,n}$ measures -- simply speaking -- the ratio of tree to 
penguin topologies. Using $SU(3)$ flavour-symmetry arguments and data on the 
CP-averaged $B^\pm\to\pi^\pm\pi^0$ branching ratio \cite{GRL}, we obtain 
$r_{\rm c,n}\sim0.2$. The parameter $q$ describes the ratio of EW penguin 
to tree contributions, and can be fixed through $SU(3)$ flavour-symmetry 
arguments, yielding $q\sim 0.7$ \cite{NR}. In order to simplify 
(\ref{Rcn-par}) and (\ref{Acn-par}), we have assumed that $q$ is a real 
parameter, as is the case in the strict $SU(3)$ limit; for generalizations, 
see \cite{BF-neutral1}. Finally, $\delta_{\rm c,n}$ is the CP-conserving
strong phase between trees and penguins. 

Consequently, the observables $R_{\rm c,n}$ and $A_0^{\rm c,n}$ depend on 
the two ``unknowns'' $\delta_{\rm c,n}$ and $\gamma$. If we vary them within 
their allowed ranges, i.e.\ $-180^\circ\leq \delta_{\rm c,n}\leq+180^\circ$ 
and $0^\circ\leq \gamma \leq180^\circ$, we obtain an allowed region in 
the $R_{\rm c,n}$--$A_0^{\rm c,n}$ plane \cite{FlMa1,FlMa2}. If the
measured values of $R_{\rm c,n}$ and $A_0^{\rm c,n}$ should lie outside
this region, we would have an immediate signal for new physics. On the
other hand, if the measurements should fall into the allowed range,
$\gamma$ and $\delta_{\rm c,n}$ could be extracted. In this case, $\gamma$
could be compared with the results of alternative strategies and
the values implied by the ``standard analysis'' of the unitarity triangle
discussed in Section~\ref{sec:intro}, whereas $\delta_{\rm c,n}$ provides 
valueable insights into hadron dynamics, thereby allowing tests of 
theoretical predictions. 

\begin{figure}[t]
\vspace*{-0.8cm}
$$\hspace*{-1.cm}
\epsfysize=0.2\textheight
\epsfxsize=0.3\textheight
\epsffile{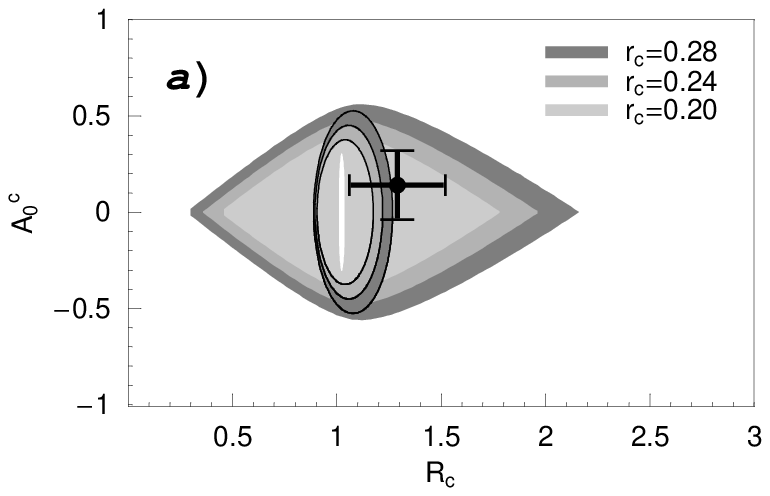} \hspace*{0.3cm}
\epsfysize=0.2\textheight
\epsfxsize=0.3\textheight
\epsffile{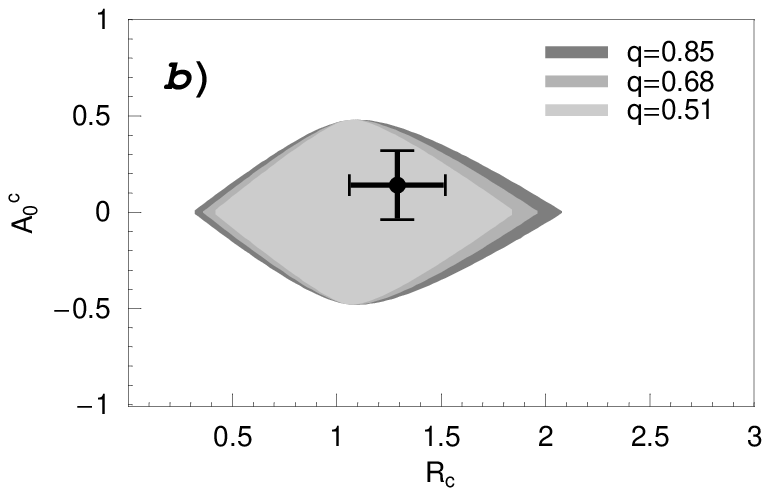}
$$
\vspace*{-0.6cm}
$$\hspace*{-1.cm}
\epsfysize=0.2\textheight
\epsfxsize=0.3\textheight
\epsffile{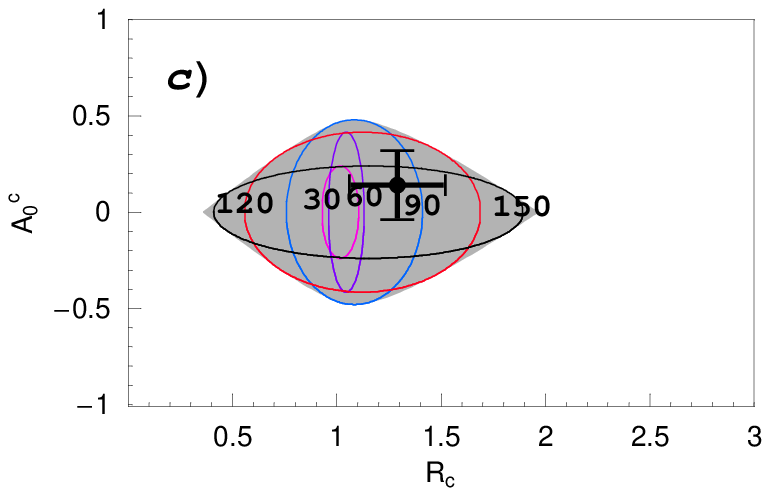} \hspace*{0.3cm}
\epsfysize=0.2\textheight
\epsfxsize=0.3\textheight
\epsffile{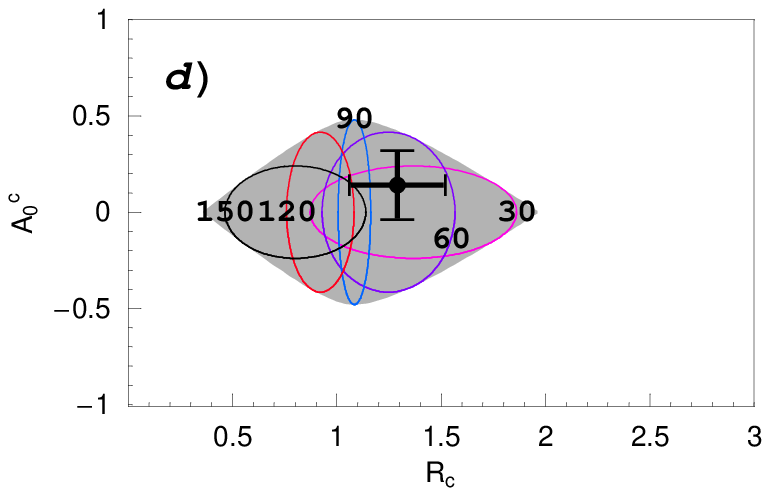}
$$
\vspace*{-0.9cm}
\caption[]{The allowed regions in the $R_{\rm c}$--$A_0^{\rm c}$ plane: (a) 
corresponds to $0.20 \leq r_{\rm c} \leq 0.28$ 
for $q=0.68$, and (b) to $0.51 \leq q \leq 0.85$ for $r_{\rm c}=0.24$;
the elliptical regions arise if we restrict $\gamma$ to the Standard-Model
range specified in (\ref{gamma-SM}). In (c) and (d), we show the contours 
for fixed values of $\gamma$ and $|\delta_{\rm c}|$, respectively 
($r_{\rm c}=0.24$, $q=0.68$).}\label{fig:BpiK-charged}
\end{figure}

In Fig.~\ref{fig:BpiK-charged}, we show the allowed regions in the 
$R_{\rm c}$--$A_0^{\rm c}$ plane for various parameter sets. The crosses 
represent the averages of the experimental results given in 
Tables~\ref{tab:BPIK-obs} and \ref{fig:BpiK-charged}. If $\gamma$ is 
constrained to 
\begin{equation}\label{gamma-SM}
50^\circ\lsim\gamma\lsim70^\circ,
\end{equation}
which corresponds to the ``Standard-Model'' range for $\gamma$ implied by the 
fits of the unitarity triangle (see (\ref{UT-Fit-ranges})), a considerably 
more restricted range arises in the $R_{\rm c}$--$A_0^{\rm c}$ plane. 
The contours in Figs.~\ref{fig:BpiK-charged} (c) and (d) allow us to read
off straightforwardly the preferred values for $\gamma$ and $\delta_{\rm c}$,
respectively, from the measured observables \cite{FlMa2}. Interestingly,
the present data seem to favour $\gamma\gsim90^\circ$ (see also
\cite{neubert-02}), which would be in conflict with (\ref{gamma-SM}). 
Moreover, they point towards $|\delta_{\rm c}|\lsim90^\circ$; factorization 
predicts $\delta_{\rm c}$ to be close to $0^\circ$ \cite{QCDF}. If future, 
more accurate data should really yield a value for $\gamma$ in the second
quadrant, the discrepancy with (\ref{gamma-SM}) may be due to new-physics
contributions to $B^0_q$--$\overline{B^0_q}$ mixing ($q\in\{d,s\}$) or the 
$B\to\pi K$ decay amplitudes. In the former case, the constraints due to 
(\ref{DMs-constr}), which rely on the Standard-Model interpretation of
$B^0_q$--$\overline{B^0_q}$ mixing, would no longer hold, so that 
$\gamma$ may actually be larger than $90^\circ$. As we shall see in 
Section~\ref{sec:Bpipi}, new data on CP violation in $B_d\to\pi^+\pi^-$
would allow us to accommodate also this picture \cite{FlMa2}. In the latter 
case, the Standard-Model expressions (\ref{Rcn-par}) and (\ref{Acn-par}) 
would receive corrections due to new physics, so that also the extracted 
value for $\gamma$ would not correspond to the Standard-Model result. More
detailed and explicit discussions of such new-physics scenarios can be 
found in \cite{BpiK-NP}.

The allowed regions and contours for the neutral $B\to\pi K$ system in 
the $R_{\rm n}$--$A_0^{\rm n}$ plane look very similar to those shown in 
Fig.~\ref{fig:BpiK-charged} \cite{FlMa2}. Unfortunately, the experimental 
situation in the neutral $B\to\pi K$ system is still rather unstable. For 
instance, in the recent Belle update \cite{bartoldus}, the central value 
for $R_{\rm n}$ has moved from 0.6 to 1.4. In the neutral strategy, another 
interesting observable is provided by the mixing-induced CP asymmetry of 
$B_d\to\pi^0K_{\rm S}$, which arises in the corresponding time-dependent 
rate asymmetry of the following kind~\cite{RF-Phys-Rep}:
\begin{equation}\label{time-asym}
\frac{\Gamma(B^0_q(t)\to f)-
\Gamma(\overline{B^0_q}(t)\to \overline{f})}{\Gamma(B^0_q(t)\to f)+
\Gamma(\overline{B^0_q}(t)\to \overline{f})}=
{\cal A}_{\rm CP}^{\rm dir}(B_q\to f)\cos(\Delta M_q t)+
{\cal A}_{\rm CP}^{\rm mix}(B_q\to f)\sin(\Delta M_q t).
\end{equation}
In the Standard Model, we expect \cite{PAPIII}
\begin{equation}
{\cal A}_{\rm CP}^{\rm mix}(B_d\to \pi^0K_{\rm S})=
{\cal A}_{\rm CP}^{\rm mix}(B_d\to J/\psi K_{\rm S}),
\end{equation}
which may well be affected by new-physics contributions, preferrably to
the $B_d\to \pi^0K_{\rm S}$ amplitudes. Concerning the extraction of
$\gamma$ from neutral $B\to\pi K$ decays discussed above, this CP asymmetry 
would allow us to take into account possible rescattering effects in an
exact manner \cite{BF-neutral1}, i.e.\ without using flavour-symmetry 
arguments.

\boldmath
\subsection{The Mixed $B\to\pi K$ System}\label{BpiK-mix}
\unboldmath
After the discussion of the charged and neutral $B\to\pi K$ systems in
Subsection~\ref{BpiK-cn}, the mixed $B\to\pi K$ system consisting of
$B^\pm\to\pi^\pm K$, $B_d\to\pi^\mp K^\pm$ can be described straightforwardly
by just making appropriate replacements of variables: first, we have
$r_{\rm c,n}\to r$, where the determination of $r$ requires the use
of factorizaton to fix a colour-allowed amplitude $T$ 
\cite{PAPIII,FM,GR,GR-BpiK-recent,QCDF}, or a measurement of 
$B_s\to\pi^\pm K^\mp$ and $U$-spin arguments \cite{GR-Uspin}. Second, we 
may set $q\to 0$, as EW penguins contribute only in colour-suppressed form 
\cite{PAPIII,QCDF}. 
We obtain then the following expressions:
\begin{eqnarray}
R&=&1-2r\cos\gamma\cos\delta+r^2\\
A_0&=&2r\sin\delta\sin\gamma,
\end{eqnarray}
which are symmetric under $\gamma\leftrightarrow\delta$. If we vary the
CP-conserving strong phase $\delta$ and the CP-violating angle $\gamma$ 
within their physical ranges, we obtain an allowed region in the 
$R$--$A_0$ plane, as shown in \cite{FlMa2}. The experimental data fall 
well into the allowed region, but do not yet allow us to draw further 
definite conclusions. At present, the situation in the charged and neutral 
$B\to\pi K$ systems appears to be more exciting.

\boldmath
\subsection{Other Recent $B\to\pi K$ Analyses}
\unboldmath
A study complementary to the one of the allowed regions in observable 
space discussed above was performed in \cite{ital-corr}, where 
the allowed regions in the $\gamma$--$\delta_{\rm c,n}$ planes implied 
by $B\to\pi K$ data were explored. Another recent $B\to\pi K$ analysis 
can be found in \cite{GR-BpiK-recent}, where the $R_{\rm (c)}$ were 
calculated for given values of $A_0^{\rm (c)}$ as functions of $\gamma$, 
and were compared with the $B$-factory data. Making more extensive use
of theory than in the flavour-symmetry strategies discussed in 
Subsections~\ref{BpiK-cn} and \ref{BpiK-mix}, several different
avenues to extract $\gamma$ from $B\to\pi K$ modes are provided by the
QCD factorization approach \cite{QCDF}, which allows also a reduction
of the theoretical uncertainties of the flavour-symmetry approaches, 
in particular a better control of $SU(3)$-breaking effects. In order
to analyse $B\to\pi K$ data, also a set of sum rules relating CP-averaged 
branching ratios and CP asymmetries of $B \to \pi K$ modes may be useful
\cite{matias}.

\boldmath
\section{Extracting $\phi_d$ from $B_d\to J/\psi K_{\rm S}$}\label{sec:phid}
\unboldmath
\boldmath
\subsection{The ``Gold-Plated'' Mode $B_d\to J/\psi K_{\rm S}$}
\unboldmath
Since penguin topologies enter in $B_d\to J/\psi K_{\rm S}$ essentially 
with the same weak phase as tree-diagram-like contributions, i.e.\ the
phase difference is doubly Cabibbo-suppressed, we obtain to a very good
approximation \cite{bisa,RF-Phys-Rep}:
\begin{equation}\label{BpsiK-CP}
{\cal A}_{\rm CP}^{\rm dir}(B_d\to J/\psi K_{\rm S})=0, \quad
{\cal A}_{\rm CP}^{\rm mix}(B_d\to J/\psi K_{\rm S})=-\sin\phi_d,
\end{equation}
where the CP-violating weak $B^0_d$--$\overline{B^0_d}$ mixing phase $\phi_d$
is given by $2\beta$ in the Standard Model. After important first steps
by the OPAL, CDF and ALEPH collaborations, the $B_d\to J/\psi K_{\rm S}$
mode (and similar decays) led eventually to the observation of CP violation 
in the $B$ system in 2001 \cite{BaBar-CP-obs,Belle-CP-obs}. The present
status of $\sin2\beta$ is given as follows:
\begin{equation}
\sin2\beta=\left\{\begin{array}{ll}
0.75\pm0.09\pm0.04&\mbox{(BaBar \cite{Babar-s2b-02})}\\
0.82\pm0.12\pm0.05&\mbox{(Belle \cite{Belle-s2b-02}),}
\end{array}\right.
\end{equation}
yielding the average of 
\begin{equation}\label{s2b-average}
\sin2\beta=0.78\pm0.08, 
\end{equation}
which agrees very well 
with the results of the CKM fits (see (\ref{UT-Fit-ranges})), 
$0.5\lsim\sin2\beta\lsim0.9$. 

In the LHC era, the experimental accuracy of the measurement of $\sin2\beta$ 
may be increased by one order of magnitude \cite{LHC-Report}. In view of
such a tremendous accuracy, it will then be important to obtain deeper
insights into the theoretical uncertainties affecting (\ref{BpsiK-CP}),
which are due to penguin contributions. A possibility to control them
is provided by the $B_s\to J/\psi K_{\rm S}$ channel \cite{BsPsiKS}. 
Moreover, also direct CP violation in $B\to J/\psi K$ modes allows us
to probe such penguin effects \cite{RF-rev96,FM-BpsiK}. So far, there are 
no experimental indications for non-vanishing CP asymmetries of this kind.

\boldmath
\subsection{Unambiguous Determination of $\phi_d$}
\unboldmath
Although the agreement between (\ref{s2b-average}) and the results of the
CKM fits is striking, it should not be forgotten that new physics may 
nevertheless hide in ${\cal A}_{\rm CP}^{\rm mix}(B_d\to J/\psi K_{\rm S})$. 
The point is that the key quantity is actually $\phi_d$, which is fixed 
through $\sin\phi_d=0.78\pm0.08$ up to a twofold ambiguity,
\begin{equation}\label{phid-det}
\phi_d=\left(51^{+8}_{-7}\right)^\circ \, \lor \,
\left(129^{+7}_{-8}\right)^\circ.
\end{equation}
Here the former solution would be in perfect agreement with the range
implied by the CKM fits, $30^\circ\lsim\phi_d\lsim70^\circ$, whereas
the latter would correspond to new physics. The two solutions can
be distinguished through a measurement of the sign of $\cos\phi_d$: 
in the case of $\cos\phi_d=+0.6>0$, we would conclude
$\phi_d=51^\circ$, whereas $\cos\phi_d=-0.6<0$ would point towards
$\phi_d=129^\circ$, i.e.\ new physics. 

There are several strategies on the market to resolve the twofold
ambiguity in the extraction of $\phi_d$ \cite{ambig}. Unfortunately,
they are rather challenging from a practical point of view. In the 
$B\to J/\psi K$ system, $\cos\phi_d$ can be extracted from the 
time-dependent angular distribution of the decay products of 
$B_d\to J/\psi[\to\ell^+\ell^-] K^\ast[\to\pi^0K_{\rm S}]$, if the sign 
of a hadronic parameter $\cos\delta$ involving a strong phase $\delta$ 
is fixed through factorization~\cite{DDFN}. Let us note that analyses
of this kind are already in progress at the $B$ factories \cite{itoh}.

\boldmath
\subsection{Aspects of New Physics in $B\to J/\psi K$}\label{BpsiK-NP}
\unboldmath
The preferred mechanism for new physics to manifest itself in CP-violating
effects in $B_d\to J/\psi K_{\rm S}$ is through $B^0_d$--$\overline{B^0_d}$
mixing, which arises in the Standard Model from the well-known box diagrams.
However, new physics may also enter at the $B\to J/\psi K$ amplitude level.
Employing estimates borrowed from effective field theory suggests that
the effects are at most of ${\cal O}(10\%)$ for a generic new-physics
scale $\Lambda_{\rm NP}$ in the TeV regime. In order to obtain the whole
picture, a set of appropriate observables can be introduced, using 
$B_d\to J/\psi K_{\rm S}$ and its charged counterpart 
$B^\pm\to J/\psi K^\pm$ \cite{FM-BpsiK}. So far, these observables do 
not yet indicate any deviation from the Standard Model. 

In the context of new-physics effects in the $B\to J/\psi K$ system, 
it is interesting to note that an upper bound on $\phi_d$ is implied by
an upper bound on $R_b\propto|V_{ub}/V_{cb}|$, as can be seen in
Fig.~\ref{fig:UT-det}. To be specific, we have 
\begin{equation}
\sin\beta_{\rm max}=R_b^{\rm max},
\end{equation}
yielding $\left(\phi_d\right)_{\rm max}^{\rm SM}\sim55^\circ$ for 
$R_b^{\rm max}\sim0.46$. As the determination of $R_b$ from semi-leptonic
tree-level decays is very robust concerning the impact of new physics,
$\phi_d\sim 129^\circ$ would require new-physics contributions to 
$B^0_d$--$\overline{B^0_d}$ mixing. As we will see in the subsequent section, 
an interesting connection between the two solutions for $\phi_d$ and 
constraints on $\gamma$ is provided by CP violation in $B_d\to\pi^+\pi^-$ 
\cite{FlMa2}. 

Concerning the search for new physics, many other promising strategies can 
now be performed in practice at the $B$ factories. An important example is 
the decay $B_d\to \phi K_{\rm S}$, where -- within the Standard Model --
mixing-induced CP violation is to a good accuracy equal to the one in 
$B_d\to J/\psi K_{\rm S}$ \cite{RF-rev96,growo,loso}, and direct CP 
violation is expected to be small. In analogy to the $B\to J/\psi K$ 
system \cite{FM-BpsiK}, also here $B^\pm\to\phi K^\pm$ decays should be 
considered as well in order to obtain the whole picture \cite{FM-BphiK}. Let 
us note that the $B$ factories have already observed these modes, and that
first measurements of ${\cal A}_{\rm CP}^{\rm mix}(B_d\to \phi K_{\rm S})$ 
were reported very recently by BaBar and Belle \cite{BphiK-exp}. In the 
future, the experimental situation will improve dramatically.

At this conference, a much more detailed discussion of $B$ physics beyond 
the Standard Model was given by David London \cite{london}. Let us now 
focus on the decay $B_d\to\pi^+\pi^-$.

\boldmath
\section{Extracting Weak Phases from $B_d\to\pi^+\pi^-$}\label{sec:Bpipi}
\unboldmath
\subsection{The Penguin Problem}
In contrast to $B_d\to J/\psi K_{\rm S}$, the relevant penguin parameter
is not doubly Cabibbo suppressed in the $B_d\to\pi^+\pi^-$ decay amplitude,
leading to the well-known ``penguin problem'' in $B_d\to\pi^+\pi^-$. If we 
had negligible penguin contributions, the corresponding CP-violating 
observables were given as follows \cite{RF-Phys-Rep}:
\begin{equation}
{\cal A}_{\rm CP}^{\rm dir}(B_d\to\pi^+\pi^-)=0, \quad
{\cal A}_{\rm CP}^{\rm mix}(B_d\to\pi^+\pi^-)=\sin(2\beta+2\gamma)=
-\sin 2\alpha,
\end{equation}
where we have also used the unitarity relation $2\beta+2\gamma=2\pi-2\alpha$. 
We observe that actually the phases $2\beta=\phi_d$ and $\gamma$ 
enter directly in the $B_d\to\pi^+\pi^-$ observables, and not $\alpha$. 
Consequently, since $\phi_d$ can be fixed straightforwardly through 
$B_d\to J/\psi K_{\rm S}$, we may use $B_d\to\pi^+\pi^-$ to probe $\gamma$. 
This is advantageous to deal with penguins and possible new-physics effects, 
as we will see below.

Measurements of the CP-violating $B_d\to\pi^+\pi^-$ observables are
already available:
\begin{equation}\label{Adir-exp}
{\cal A}_{\rm CP}^{\rm dir}(B_d\to\pi^+\pi^-)=\left\{
\begin{array}{ll}
-0.02\pm0.29\pm0.07 & \mbox{(BaBar \cite{BaBar-Bpipi-new})}\\
-0.94^{+0.31}_{-0.25}\pm0.09 & \mbox{(Belle \cite{Belle-Bpipi})}
\end{array}
\right.
\end{equation}
\begin{equation}\label{Amix-exp}
{\cal A}_{\rm CP}^{\rm mix}(B_d\to\pi^+\pi^-)=\left\{
\begin{array}{ll}
0.01\pm0.37\pm0.07& \mbox{(BaBar \cite{BaBar-Bpipi-new})}\\
1.21^{+0.27+0.13}_{-0.38-0.16} & \mbox{(Belle \cite{Belle-Bpipi}).}
\end{array}
\right.
\end{equation}
Unfortunately, the BaBar and Belle results are not fully consistent with
each other. This discrepancy will hopefully be resolved soon. Forming 
nevertheless the na\"\i ve averages of the numbers in (\ref{Adir-exp}) and 
(\ref{Amix-exp}) yields
\begin{equation}\label{Bpipi-CP-averages}
{\cal A}_{\rm CP}^{\rm dir}(B_d\to\pi^+\pi^-)=-0.48\pm0.21,\quad
{\cal A}_{\rm CP}^{\rm mix}(B_d\to\pi^+\pi^-)=0.61\pm0.26.
\end{equation}
Direct CP violation at this level would require large penguin contributions 
with large CP-conserving strong phases. A significant impact of penguins on 
$B_d\to\pi^+\pi^-$ is also indicated by data on $B\to\pi K,\pi\pi$ decays, 
as well as by theoretical considerations \cite{QCDF,PQCD,ital-charm}. 
Consequently, it is already evident that the penguin contributions to
$B_d\to\pi^+\pi^-$ {\it cannot} be neglected.

\subsection{Possible Solutions of the Penguin Problem}
There are many approaches to deal with the penguin problem in the extraction 
of weak phases from CP violation in $B_d\to\pi^+\pi^-$; the best known is an 
isospin analysis of the $B\to\pi\pi$ system \cite{GL}, yielding $\alpha$.
Unfortunately, this approach is very difficult in practice, as it requires
a measurement of the $B_d\to\pi^0\pi^0$ branching ratio. However,
useful bounds may already be obtained from experimental constraints on
this branching ratio \cite{alpha-bounds,charles}.

Alternatively, we may employ the CKM unitarity to express 
${\cal A}_{\rm CP}^{\rm mix}(B_d\to\pi^+\pi^-)$ in terms of $\alpha$ and 
hadronic parameters. Using ${\cal A}_{\rm CP}^{\rm dir}(B_d\to\pi^+\pi^-)$, 
a strong phase can be eliminated, allowing us to determine $\alpha$ as
a function of a hadronic parameter $|p/t|$, which is, however, problematic
to be determined reliably \cite{QCDF,ital-charm,charles,FM1,LSS,LX}.

A different parametrization of $B_d\to\pi^+\pi^-$, involving a parameter 
$de^{i\theta}\equiv-T/P$ and $\phi_d=2\beta$, is employed in 
\cite{GR-Bpipi1}, where, moreover, $\alpha+\beta+\gamma=180^\circ$ is used 
to eliminate $\gamma$, and $\beta$ is fixed through the Standard-Model 
solution $\sim 26^\circ$ implied by ${\cal A}_{\rm CP}^{\rm mix}
(B_d\to J/\psi K_{\rm S})$. Provided $d$ is known, $\alpha$ can be extracted
from the CP-violating $B_d\to\pi^+\pi^-$ observables. To this end, $SU(3)$ 
flavour-symmetry arguments and plausible dynamical assumptions are used
to fix $|P|$ through the CP-averaged $B^\pm\to\pi^\pm K$ branching ratio. 
On the other hand, $|T|$ is estimated with the help of factorization and 
data on $B\to\pi\ell\nu$. Refinements of this approach were presented in 
\cite{GR-Bpipi2}.

Another strategy to deal with penguins in $B_d\to\pi^+\pi^-$ is offered 
by $B_s\to K^+K^-$. Using the $U$-spin symmetry of strong interactions, 
$\phi_d$ and $\gamma$ can be extracted from the corresponding CP-violating 
observables \cite{RF-BsKK}. In the following discussion, we shall put a
particular emphasis on a variant for the $e^+e^-$ $B$-factories, where 
$B_s\to K^+K^-$ is replaced by $B_d\to\pi^\mp K^\pm$ \cite{U-variant}. 
We may then already use the $B$-factory data to explore allowed regions in 
observable space and to extract weak phases and hadronic parameters 
\cite{FlMa2}.

\boldmath
\subsection{The $B_d\to\pi^+\pi^-$, $B_s\to K^+K^-$ 
System}\label{subsec:Bpipi-BsKK}
\unboldmath
As can be seen from the corresponding Feynman diagrams, $B_s\to K^+K^-$ 
is related to $B_d\to\pi^+\pi^-$ through an interchange of all down and 
strange quarks. The decay amplitudes read as follows \cite{RF-BsKK}:
\begin{equation}\label{Bpipi-ampl}
A(B_d^0\to\pi^+\pi^-)={\cal C}\left[e^{i\gamma}-d e^{i\theta}\right]
\end{equation}
\begin{equation}\label{BsKK-ampl}
A(B_s^0\to K^+K^-)=\left(\frac{\lambda}{1-\lambda^2/2}\right){\cal C}'
\left[e^{i\gamma}+\left(\frac{1-\lambda^2}{\lambda^2}\right)
d'e^{i\theta'}\right],
\end{equation}
where the CP-conserving strong amplitudes $d e^{i\theta}$ and 
$d'e^{i\theta'}$ measure, sloppily speaking, ratios of penguin to tree 
amplitudes in $B_d^0\to\pi^+\pi^-$ and $B_s^0\to K^+K^-$, respectively.
Using these general parametrizations, we obtain expressions for the 
direct and mixing-induced CP asymmetries of the following kind:
\begin{equation}\label{Bpipi-obs}
{\cal A}_{\rm CP}^{\rm dir}(B_d\to\pi^+\pi^-)=
\mbox{fct}(d,\theta,\gamma), \quad
{\cal A}_{\rm CP}^{\rm mix}(B_d\to\pi^+\pi^-)=
\mbox{fct}(d,\theta,\gamma,\phi_d)
\end{equation}
\begin{equation}\label{BsKK-obs}
{\cal A}_{\rm CP}^{\rm dir}(B_s\to K^+K^-)=
\mbox{fct}(d',\theta',\gamma), \quad
{\cal A}_{\rm CP}^{\rm mix}(B_s\to K^+K^-)=
\mbox{fct}(d',\theta',\gamma,\phi_s\approx0).
\end{equation}

Consequently, we have four observables at our disposal, depending on six 
``unknowns''. However, since $B_d\to\pi^+\pi^-$ and $B_s\to K^+K^-$ are 
related to each other by interchanging all down and strange quarks, the 
$U$-spin flavour symmetry of strong interactions implies
\begin{equation}\label{U-spin-rel}
d'e^{i\theta'}=d\,e^{i\theta}.
\end{equation}
Using this relation, the four observables in (\ref{Bpipi-obs}) and
(\ref{BsKK-obs}) depend on the four quantities $d$, $\theta$, 
$\phi_d$ and $\gamma$, which can hence be determined \cite{RF-BsKK}. 
The theoretical accuracy is
only limited by the $U$-spin symmetry, as no dynamical assumptions about
rescattering processes have to be made. Theoretical considerations give us
confidence into (\ref{U-spin-rel}), as it does not receive $U$-spin-breaking 
corrections within factorization \cite{RF-BsKK}. Moreover, we may also obtain 
experimental insights into $U$-spin breaking \cite{RF-BsKK,gronau-U-spin}. 

The $U$-spin arguments can be minimized, if the $B^0_d$--$\overline{B^0_d}$ 
mixing phase $\phi_d$, which can be fixed through 
$B_d\to J/\psi K_{\rm S}$, is used as an input. The observables 
${\cal A}_{\rm CP}^{\rm dir}(B_d\to\pi^+\pi^-)$ and
${\cal A}_{\rm CP}^{\rm mix}(B_d\to\pi^+\pi^-)$ allow us then to 
eliminate the strong phase $\theta$ and to determine $d$ as a function of
$\gamma$. Analogously, ${\cal A}_{\rm CP}^{\rm dir}(B_s\to K^+K^-)$ and 
${\cal A}_{\rm CP}^{\rm mix}(B_s\to K^+K^-)$ allow us to eliminate 
the strong phase $\theta'$ and to determine $d'$ as a function of
$\gamma$. The corresponding contours in the $\gamma$--$d$
and $\gamma$--$d'$ planes can be fixed in a {\it theoretically clean} way.
Using now the $U$-spin relation $d'=d$, these contours allow the 
determination both of the CKM angle $\gamma$ and of the hadronic quantities 
$d$, $\theta$, $\theta'$; for a detailed illustration, see \cite{RF-BsKK}.
This approach is very promising for run II of the Tevatron and the experiments
of the LHC era, where experimental accuracies for $\gamma$ of 
${\cal O}(10^\circ)$ \cite{Paulini,TEV-II} and ${\cal O}(1^\circ)$ 
\cite{LHC-Report} may be achieved, respectively. It should be emphasized 
that not only $\gamma$, but also the hadronic parameters $d$, $\theta$, 
$\theta'$ are of particular interest, as they can be compared with theoretical 
predictions, thereby allowing important insights into hadron dynamics. For 
other recently developed $U$-spin strategies, the reader is referred to 
\cite{GR-Uspin,BsPsiKS,skands}.

\boldmath
\subsection{The $B_d\to\pi^+\pi^-$, $B_d\to \pi^\mp K^\pm$ 
System}\label{subsec:Bpipi-analysis}
\unboldmath
Since $B_s\to K^+K^-$ is not accessible at the $e^+e^-$ $B$-factories 
operating at $\Upsilon(4S)$, data are not yet available.
However, as can be seen by looking at the corresponding Feynman diagrams,
$B_s\to K^+K^-$ is related to $B_d\to\pi^\mp K^\pm$ through an interchange 
of spectator quarks. Consequently, we have
\begin{equation}\label{CP-rel}
{\cal A}_{\rm CP}^{\rm dir}(B_s\to K^+K^-)\approx{\cal A}_{\rm CP}^{\rm dir}
(B_d\to\pi^\mp K^\pm)
\end{equation}
\begin{equation}\label{BR-rel}
\mbox{BR}(B_s\to K^+K^-)
\approx\mbox{BR}(B_d\to\pi^\mp K^\pm)\,\frac{\tau_{B_s}}{\tau_{B_d}}.
\end{equation}
For the following considerations, the quantity 
\begin{equation}\label{H-def}
H\equiv\frac{1}{\epsilon}\left|
\frac{{\cal C}'}{{\cal C}}\right|^2
\left[\frac{\mbox{BR}(B_d\to\pi^+\pi^-)}{\mbox{BR}(B_s\to K^+K^-)}\right]
\end{equation}
is particularly useful \cite{U-variant}, where 
\begin{equation}
\epsilon\equiv\frac{\lambda^2}{1-\lambda^2}, 
\end{equation}
and ${\cal C}$ and ${\cal C}'$ are the normalization factors introduced in 
(\ref{Bpipi-ampl}) and (\ref{BsKK-ampl}), respectively. Using (\ref{CP-rel}) 
and (\ref{BR-rel}), as well as factorization to estimate $U$-spin-breaking 
corrections to 
\begin{equation}\label{C-rel}
{\cal C}'={\cal C}, 
\end{equation}
$H$ can be determined from the $B$-factory data as follows:
\begin{equation}\label{H-det}
H\approx\frac{1}{\epsilon}\left(\frac{f_K}{f_\pi}\right)^2
\left[\frac{\mbox{BR}(B_d\to\pi^+\pi^-)}{\mbox{BR}(B_d\to\pi^\mp K^\pm)}
\right]=
\left\{\begin{array}{ll}
7.3\pm2.9 & \mbox{(CLEO)}\\
8.8\pm1.5 & \mbox{(BaBar)}\\
6.8\pm1.7 & \mbox{(Belle).}
\end{array}\right.
\end{equation}

\begin{figure}
\vspace*{-0.7truecm}
\centerline{{
\vspace*{-0.2truecm}
\epsfysize=4.8truecm
{\epsffile{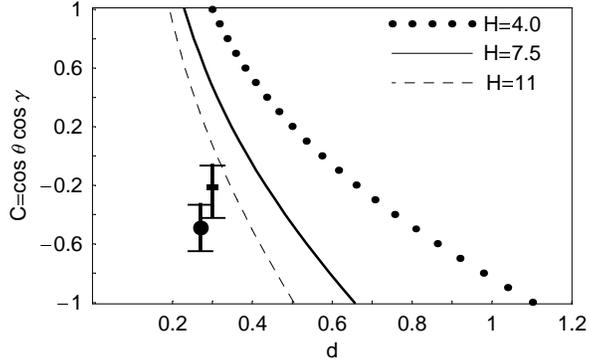}}}}
\caption{The dependence of $C\equiv\cos\theta\cos\gamma$ on $d$ for values of 
$H$ consistent with (\ref{H-det}). The ``circle'' and ``square'' with error 
bars represent the predictions of the QCD factorization \cite{QCDF} and 
PQCD \cite{PQCD-pred} approaches, respectively, for the Standard-Model 
range (\ref{gamma-SM}) of $\gamma$.}\label{fig:C-d}
\end{figure}

If we use the $U$-spin relation (\ref{U-spin-rel}) and the amplitude 
parametrizations in (\ref{Bpipi-ampl}) and (\ref{BsKK-ampl}), we obtain
\begin{equation}\label{H-expr}
H=\frac{1-2 d\cos\theta\cos\gamma+d^2}{\epsilon^2+
2\epsilon d\cos\theta\cos\gamma+d^2}.
\end{equation}
Consequently, $H$ allows us to determine $C\equiv\cos\theta\cos\gamma$ 
as a function of $d$, as shown in Fig.~\ref{fig:C-d}. We observe that
the rather restricted range $0.2\lsim d\lsim 1$ is implied by the data, 
demonstrating the importance of penguins in $B_d\to\pi^+\pi^-$. Moreover, 
the experimental curves are not in favour of a Standard-Model interpretation 
of the theoretical predictions for $de^{i\theta}$ within the QCD factorization
\cite{QCDF} and PQCD \cite{PQCD-pred} approaches. Interestingly, agreement
could be achieved for $\gamma>90^\circ$, as the circle and square in
Fig.~\ref{fig:C-d}, calculated for $\gamma=60^\circ$, would then move to 
positive values of $C$ \cite{FlMa2,U-variant}.

Let us now come back to the decay $B_d\to\pi^+\pi^-$ and its CP-violating
observables, as parametrized in (\ref{Bpipi-obs}). As we have already noted,
$\phi_d$ entering ${\cal A}_{\rm CP}^{\rm mix}(B_d\to\pi^+\pi^-)$
can be fixed through ${\cal A}_{\rm CP}^{\rm mix}(B_d\to J/\psi K_{\rm S})$,
yielding the twofold solution in (\ref{phid-det}). We may now employ 
$H$ as an additional observable to deal with the penguins. Applying
(\ref{U-spin-rel}), we obtain $H=\mbox{fct}(d,\theta,\gamma)$, as given
explicitly in (\ref{H-expr}). We may then eliminate the penguin parameter
$d$ in (\ref{Bpipi-obs}) through $H$. If we vary the remaining parameters
$\theta$ and $\gamma$ within their physical ranges, i.e.\ 
$-180^\circ\leq \theta\leq+180^\circ$ and 
$0^\circ\leq \gamma \leq180^\circ$, we obtain an allowed region in the
${\cal A}_{\rm CP}^{\rm dir}(B_d\to\pi^+\pi^-)$--${\cal A}_{\rm CP}^{\rm 
mix}(B_d\to\pi^+\pi^-)$ plane. If the measured observables should lie within 
this region, we may extract $\gamma$, $\theta$ and $d$ 
\cite{FlMa2,U-variant}. The value for $\gamma$ may then be compared with 
results of other strategies or the ``Standard-Model'' range (\ref{gamma-SM}), 
whereas the hadronic parameters $\theta$ and $d$ allow us to test theoretical 
predictions.

\begin{figure}
\vspace*{-0.5cm}
$$\hspace*{-1.cm}
\epsfysize=0.2\textheight
\epsfxsize=0.3\textheight
\epsffile{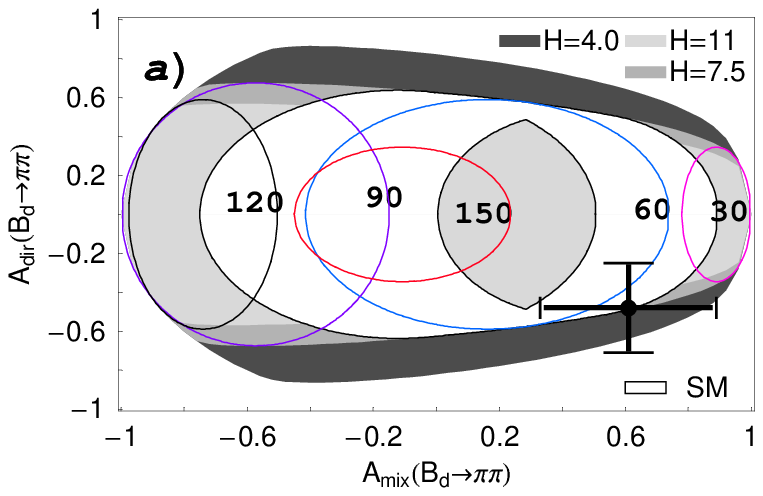} \hspace*{0.3cm}
\epsfysize=0.2\textheight
\epsfxsize=0.3\textheight
\epsffile{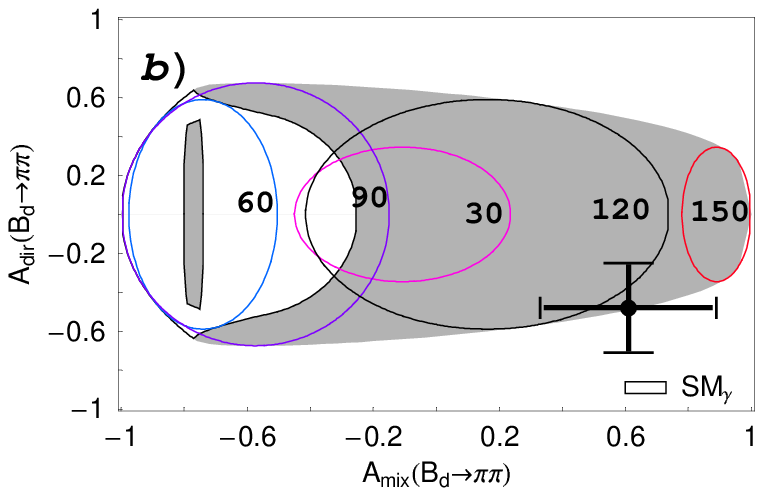}
$$
\vspace*{-0.9cm}
\caption[]{Allowed region in the 
${\cal A}_{\rm CP}^{\rm mix}(B_d\to\pi^+\pi^-)$--${\cal A}_{\rm CP}^{\rm
dir}(B_d\to\pi^+\pi^-)$ plane for (a) $\phi_d=51^\circ$ and various values
of $H$, and (b) $\phi_d=129^\circ$ ($H=7.5$). The SM regions arise if we 
restrict $\gamma$ to (\ref{gamma-SM}) ($H=7.5$). Contours representing
fixed values of $\gamma$ are also included.}\label{fig:AdAmpipi}
\end{figure}

\begin{figure}
$$\hspace*{-1.cm}
\epsfysize=0.2\textheight
\epsfxsize=0.3\textheight
\epsffile{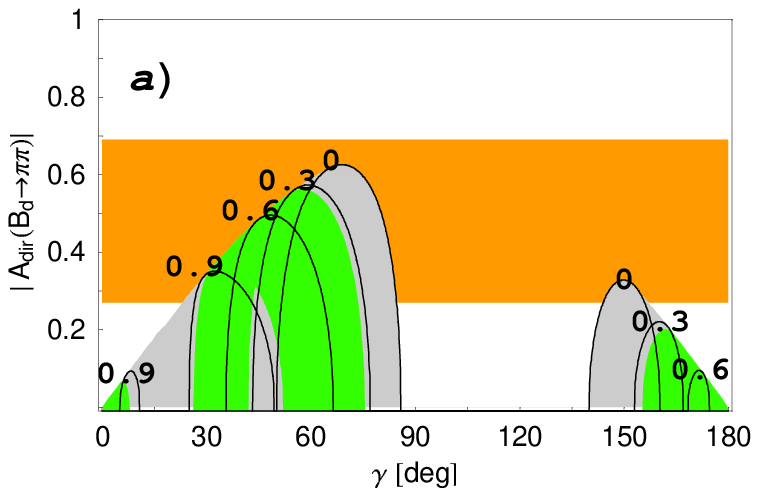} \hspace*{0.3cm}
\epsfysize=0.2\textheight
\epsfxsize=0.3\textheight
\epsffile{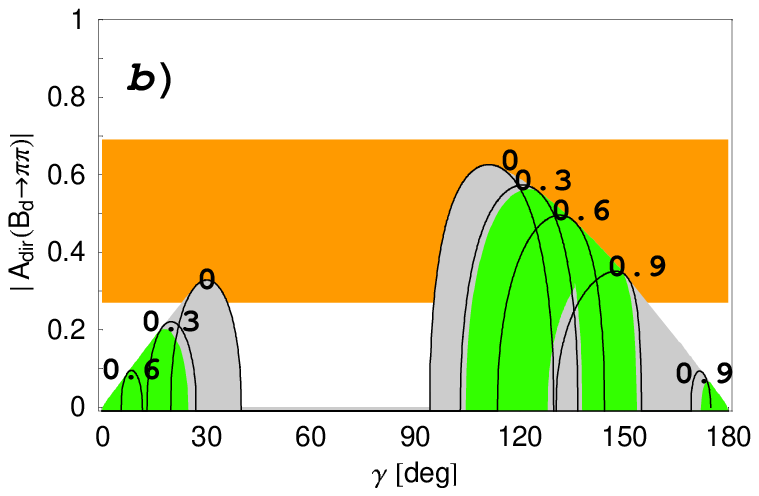}
$$
\vspace*{-0.5cm}
$$\hspace*{-1.cm}
\epsfysize=0.2\textheight
\epsfxsize=0.3\textheight
\epsffile{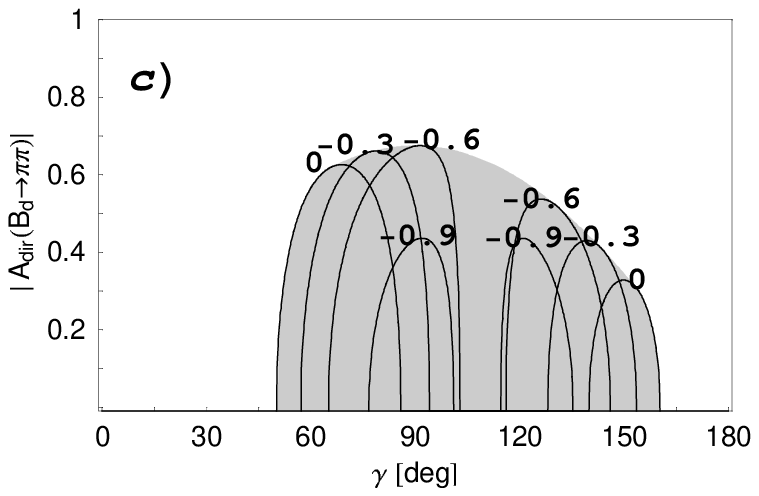} \hspace*{0.3cm}
\epsfysize=0.2\textheight
\epsfxsize=0.3\textheight
 \epsffile{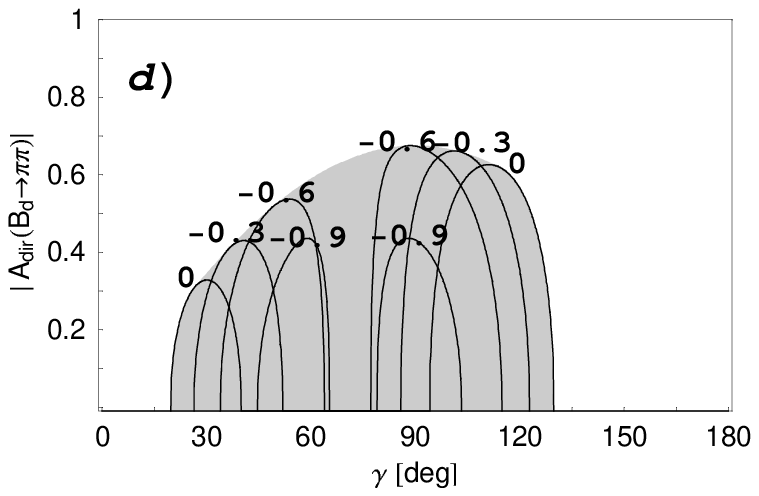}
$$
\vspace*{-0.9cm}
\caption[]{$|{\cal A}_{\rm CP}^{\rm dir}(B_d\to\pi^+\pi^-)|$ as a function
of $\gamma$ for various values of 
${\cal A}_{\rm CP}^{\rm mix}(B_d\to\pi^+\pi^-)$ ($H=7.5$). 
In (a) and (b), $\phi_d=51^\circ$ and $\phi_d=129^\circ$ were chosen,
respectively. The shaded region arises from a variation of 
${\cal A}_{\rm CP}^{\rm mix}(B_d\to\pi^+\pi^-)$ within $[0,+1]$. The
corresponding plots for negative  
${\cal A}_{\rm CP}^{\rm mix}(B_d\to\pi^+\pi^-)$ are shown in (c) and
(d) for $\phi_d=51^\circ$ and $\phi_d=129^\circ$, respectively. The 
bands arising from the experimental averages in (\ref{Bpipi-CP-averages})
are also included.}\label{fig:gam-Add}
\end{figure}

In Fig.~\ref{fig:AdAmpipi}, we show the allowed regions in the
${\cal A}_{\rm CP}^{\rm dir}(B_d\to\pi^+\pi^-)$--${\cal A}_{\rm CP}^{\rm 
mix}(B_d\to\pi^+\pi^-)$ plane for the two solutions of $\phi_d$ and
various values of $H$ \cite{FlMa2}, as well as the contours arising for
fixed values of $\gamma$. We observe that the experimental averages, 
represented by the crosses, overlap nicely with the SM region for 
$\phi_d=51^\circ$, and point towards $\gamma\sim50^\circ$. In this case, 
not only $\gamma$ would be in accordance with the results 
of the CKM fits described in Section~\ref{sec:intro}, but also the 
$B^0_d$--$\overline{B^0_d}$ mixing phase $\phi_d$. On the other hand, for 
$\phi_d=129^\circ$, the experimental values favour $\gamma\sim130^\circ$, 
and have essentially no overlap with the SM region. Since a value of 
$\phi_d=129^\circ$ would require CP-violating new-physics contributions 
to $B^0_d$--$\overline{B^0_d}$ mixing, as we have seen in 
Subsection~\ref{BpsiK-NP}, also the $\gamma$ range in (\ref{gamma-SM}) may 
no longer hold in this case, as it relies on a Standard-Model interpretation 
of the experimental information on $B^0_{d,s}$--$\overline{B^0_{d,s}}$ mixing.
In particular, also values for $\gamma$ larger than $90^\circ$ could then 
in principle be accommodated. 

In order to put these observations on a more quantitative basis, we show in 
Fig.~\ref{fig:gam-Add} the dependence of 
$|{\cal A}_{\rm CP}^{\rm dir}(B_d\to\pi^+\pi^-)|$ on 
$\gamma$ for given values of ${\cal A}_{\rm CP}^{\rm mix}(B_d\to\pi^+\pi^-)$
\cite{FlMa2}. An interesting difference arises, if we consider positive and 
negative values of the mixing-induced CP asymmetry. In the former case, 
we find that the solution $\phi_d=51^\circ$ being in agreement with
the CKM fits allows us to accommodate conveniently the Standard-Model 
range (\ref{gamma-SM}), whereas we obtain a gap around $\gamma\sim 60^\circ$
for $\phi_d=129^\circ$. On the other hand, if we consider negative values 
of ${\cal A}_{\rm CP}^{\rm mix}(B_d\to\pi^+\pi^-)$, both solutions for
$\phi_d$ could accommodate (\ref{gamma-SM}), and the situation would not 
be as exciting as for a positive value of 
${\cal A}_{\rm CP}^{\rm mix}(B_d\to\pi^+\pi^-)$. Interestingly, a positive 
value is now favoured by the data. Taking into account the experimental 
averages given in (\ref{Bpipi-CP-averages}), we obtain the bands in 
Figs.~\ref{fig:gam-Add} (a) and (b), yielding the following solutions for 
$\gamma$:
\begin{equation}\label{gam-res}
28^\circ\lsim\gamma\lsim74^\circ \, (\phi_d=51^\circ), \quad
106^\circ\lsim\gamma\lsim152^\circ \, (\phi_d=129^\circ).
\end{equation}
In the future, the experimental uncertainties will be reduced considerably, 
i.e.\ the experimental bands will become much narrower, thereby providing 
significantly more stringent results for $\gamma$, as well as the hadronic
parameters. 

Let us now turn to the theoretical uncertainties affecting this approach. In 
the determination of $H$ through (\ref{H-det}), corrections to the 
$U$-spin relation (\ref{C-rel}) are taken into account through factorization, 
leading to the $f_K/f_{\pi}$ factor. Moreover, in order to replace 
$\mbox{BR}(B_s\to K^+K^-)$ through $\mbox{BR}(B_d\to \pi^\mp K^\pm)$, $SU(3)$ 
arguments and plausible dynamical assumptions have to be used. Once 
measurements of the direct and mixing-induced CP asymmetries of 
$B_s\to K^+K^-$ are available, $H$ can be circumvented, as we have seen in 
Subsection~\ref{subsec:Bpipi-BsKK}. The second kind of uncertainties enters 
through (\ref{U-spin-rel}). As we have already noted, this relation is 
not affected by $U$-spin-breaking corrections within factorization 
\cite{RF-BsKK}, since the relevant decay constants and form factors cancel 
in the corresponding ratios of decay amplitudes. The impact of 
non-factorizable effects can be described by 
\begin{equation}
\xi_d\equiv d'/d, \quad \Delta\theta\equiv\theta'-\theta.
\end{equation}
In \cite{FlMa2}, formulae are given to take into account these parameters
in an exact manner, allowing us to explore their impact. The dominant effects 
are due to $\xi_d$, whereas $\Delta\theta$ plays a minor r\^ole. A detailed
analysis of the sensitivity of the plots shown in Figs.~\ref{fig:AdAmpipi} 
and \ref{fig:gam-Add} on $\xi_d$ was performed in \cite{FlMa2}. Concerning 
the ranges for $\gamma$ given in (\ref{gam-res}), the impact of a variation 
of $\xi_d$ within $[0.8,1.2]$ is rather moderate, yielding
\begin{equation}\label{gam-res-xi}
(28\pm2)^\circ\lsim\gamma\lsim(74\pm6)^\circ \, (\phi_d=51^\circ), \quad
(106\pm6)^\circ\lsim\gamma\lsim(152\pm2)^\circ \, (\phi_d=129^\circ).
\end{equation}

Let us finally make a few comments on factorization. Here we have
$\left.\theta\right|_{\rm fact}\sim180^\circ$, implying 
${\cal A}_{\rm CP}^{\rm dir}(B_d\to\pi^+\pi^-)\sim 0$. The Belle data
are not in accordance with such a picture, favouring large penguin effects
and a large strong phase $\theta$, where the sign of 
${\cal A}_{\rm CP}^{\rm dir}(B_d\to\pi^+\pi^-)$ implies 
$0^\circ<\theta<180^\circ$. On the other hand, the BaBar data are still
consistent with factorization. Interestingly, the expressions for $H$ and
${\cal A}_{\rm CP}^{\rm mix}(B_d\to\pi^+\pi^-)$ depend only on $\cos\theta$
\cite{FlMa2}. In contrast to 
$\sin\theta$, $\cos\theta$ is not very sensitive to deviations of $\theta$ 
from $\left.\theta\right|_{\rm fact}\sim180^\circ$. If we introduce
$c\equiv-\cos\theta$ and $c'\equiv-\cos\theta'$, $H$ allows us to calculate 
$d$ and ${\cal A}_{\rm CP}^{\rm mix}(B_d\to\pi^+\pi^-)$ as functions of 
$\gamma$. Using then the experimental value of the mixing-induced
CP asymmetry, $\gamma$ and $d$ can be extracted. This determination is 
interestingly very stable with respect to $\theta,\theta'\not=180^\circ$.
In the case of $H=7.5$ and ${\cal A}_{\rm CP}^{\rm mix}(B_d\to\pi^+\pi^-)
\sim 0$, we obtain the following solutions \cite{FlMa2}:
\begin{equation}\label{gamma-fact}
\gamma\sim86^\circ \lor 160^\circ \,\, (\phi_d=51^\circ), \quad
\gamma\sim40^\circ \lor 130^\circ \,\, (\phi_d=129^\circ),
\end{equation}
with hadronic parameters
\begin{equation}\label{D-fact-ranges}
d\sim0.4 \lor 0.2 \,\, (\phi_d=51^\circ), \quad
d\sim0.6 \lor 0.3 \,\, (\phi_d=129^\circ).
\end{equation}
Since theoretical estimates prefer $d\sim0.3$, as we have seen above,
the solutions for $\gamma$ larger than $90^\circ$ would be favoured. This is
in accordance with the situation in Fig.~\ref{fig:C-d}.

\begin{figure}[t]
\vspace*{-0.4cm}
$$\hspace*{-1.cm}
\epsfysize=0.2\textheight
\epsfxsize=0.3\textheight
\epsffile{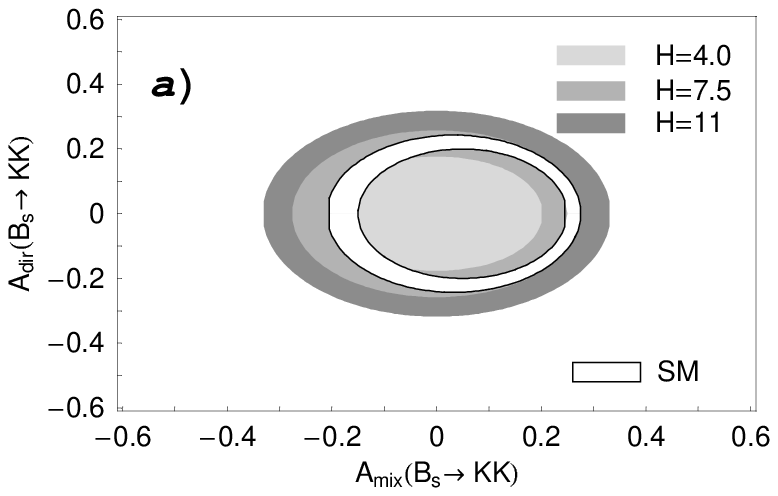} \hspace*{0.3cm}
\epsfysize=0.2\textheight
\epsfxsize=0.3\textheight
 \epsffile{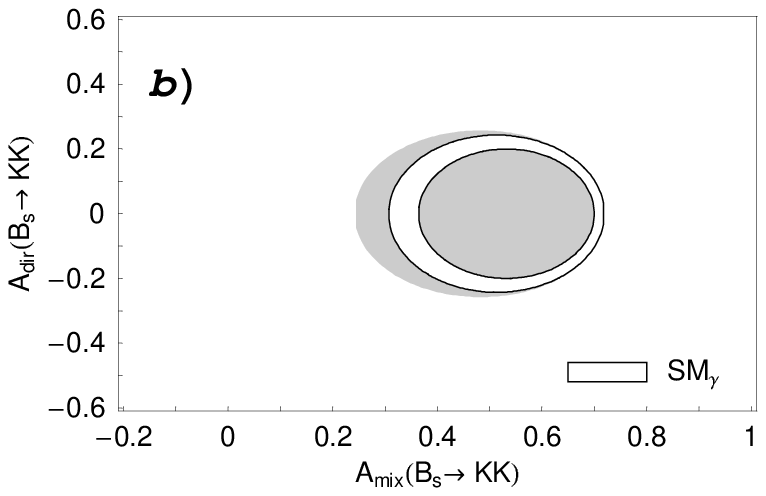}
$$
\vspace*{-0.9cm}
\caption[]{Allowed region in the
${\cal A}_{\rm CP}^{\rm mix}(B_s\to K^+K^-)$--${\cal A}_{\rm CP}^{\rm
dir}(B_s\to K^+K^-)$ plane for (a) $\phi_s=0^\circ$ and various values of 
$H$, and (b) $\phi_s^{\rm NP}=30^\circ$ ($H=7.5$). The SM regions arise if 
$\gamma$ is restricted to (\ref{gamma-SM}) ($H=7.5$).}\label{fig:Ams-Ads}
\end{figure}

\boldmath
\subsection{Allowed Regions in $B_s\to K^+K^-$ Observable Space}
\unboldmath
In analogy to the analysis of the $B_d\to\pi^+\pi^-$ mode discussed in 
Subsection~\ref{subsec:Bpipi-analysis}, we may also use $H$ to eliminate 
$d'$ in ${\cal A}_{\rm CP}^{\rm dir}(B_s\to K^+K^-)$ and
${\cal A}_{\rm CP}^{\rm mix}(B_s\to K^+K^-)$. If we vary then
$\theta'$ and $\gamma$ within their physical ranges, i.e.\
$-180^\circ\leq \theta'\leq+180^\circ$ and $0^\circ\leq \gamma \leq180^\circ$,
we obtain an allowed region in the ${\cal A}_{\rm CP}^{\rm mix}(B_s\to 
K^+K^-)$--${\cal A}_{\rm CP}^{\rm dir}(B_s\to K^+K^-)$ plane \cite{FlMa2},
as shown in Fig.~\ref{fig:Ams-Ads}. There also the impact of a
non-vanishing value of $\phi_s$, which may be due to new-physics contributions
to $B^0_s$--$\overline{B^0_s}$ mixing, is illustrated. If we constrain 
$\gamma$ to (\ref{gamma-SM}), even more restricted regions arise. The
allowed regions are remarkably stable with respect to variations of 
$\xi_d$ and $\Delta\theta$ \cite{FlMa2}, and represent a narrow target 
range for run II of the Tevatron and the experiments of the LHC era, in 
particular LHCb and BTeV.

\section{Conclusions and Outlook}\label{sec:concl}
Due to the efforts of the BaBar and Belle collaborations, CP violation
could recently be established in the $B$ system through the ``gold-plated''
mode $B_d\to J/\psi K_{\rm S}$, thereby opening a new era in the 
exploration of CP-violating phenomena. The world average $0.78\pm0.08$
of $\sin\phi_d$ determined through $B_d\to J/\psi K_{\rm S}$ and similar
modes agrees now well with the Standard Model, but leaves a twofold 
solution for $\phi_d$, given by $\phi_d=(51^{+8}_{-7})^\circ \lor
(129^{+7}_{-8})^\circ$. The former solution is in accordance with 
the picture of the Standard Model, whereas the latter would point towards 
new-physics contributions to $B^0_d$--$\overline{B^0_d}$ mixing. As we 
have seen, it is an important issue to resolve this ambiguity directly, 
also in view of an analysis of CP violation in $B_d\to\pi^+\pi^-$.
Other exciting results are expected soon. An important example is the decay
$B_d\to\phi K_{\rm S}$, where the Standard Model predicts with good 
accuracy small direct CP violation and 
${\cal A}_{\rm CP}^{\rm mix}(B_d\to \phi K_{\rm S})=
{\cal A}_{\rm CP}^{\rm mix}(B_d\to J/\psi K_{\rm S})$.

Another key element in the phenomenology of $B$ physics is given by
$B\to\pi K$ decays, complemented by data on $B\to\pi\pi$. These modes can
be described efficiently through allowed ranges in observable space, allowing
a straightforward comparison with experiment, and play an important r\^ole
to obtain information on $\gamma$. Interestingly, the data on CP-averaged
$B\to\pi K$ branching ratios seem to favour $\gamma>90^\circ$, which would
be in conflict with the Standard-Model interpretation of lower bounds on
$\Delta M_s/\Delta M_d$, pointing towards $\gamma<90^\circ$. It is still 
too early to draw definite conclusions, but the picture is expected to improve 
significantly in the future. 

Recent experimental results indicate large penguin effects in the
$B_d\to\pi^+\pi^-$ mode. The question arises now how direct and mixing-induced
CP violation in this channel can be translated -- in an experimentally
feasible way -- into results for angles of the unitarity triangle. Here
we have discussed one possiblility to achieve this goal in more detail,
using $B_d\to\pi^\mp K^\pm$ to control the penguin effects. Following these
lines, we obtain $28^\circ\lsim\gamma\lsim74^\circ$ ($\phi_d\sim51^\circ$)
$\lor$  $106^\circ\lsim\gamma\lsim152^\circ$ ($\phi_d\sim129^\circ$), where
the former solution would be in agreement with the Standard-Model picture. 
Unfortunately, the present BaBar and Belle results on CP violation in 
$B_d\to\pi^+\pi^-$ are not fully consistent with each other. This issue
will hopefully be settled soon. Once measurements of the CP-violating 
observables of $B_s\to K^+K^-$ are available, a more elegant determination 
of $\gamma$ is possible, which is also cleaner from a theoretical point of
view. Within the Standard Model, the presently available data on the ratio 
of the CP-averaged $B_d\to\pi^\mp K^\pm$ and $B_d\to\pi^+\pi^-$ branching 
ratios imply already a very constrained target region in $B_s\to K^+K^-$
observable space for run II of the Tevatron and the experiments of the LHC 
era. It will be exciting to see whether these experiments will actually hit 
the Standard-Model region.

\section*{Acknowledgements}
I am very grateful to the organizers for inviting me to this inspiring
conference, and would like to thank Joaquim Matias for a pleasant 
collaboration on many of the recent results presented above.



\end{document}